\documentstyle[12pt]{article}
\begin{document}

\thispagestyle{empty}
\newcommand{\beq}{\begin{equation}}
\newcommand{\eeq}{\end{equation}}
\newcommand{\RA}{\rightarrow}
\baselineskip=0.6cm
\begin{flushright}{.}\end{flushright}
\begin{flushright}{TIT/HEP-265}\\
{August 1994}\end{flushright}
\begin{center}
\vspace{15mm}

{\large\bf The heat kernel for deformed spheres}\\

\bigskip
\vspace{0.3in}

{N.~Shtykov}\\
\medskip{\it Department of Physics , Tokyo Institute of
Technology},\\
{\it Oh-Okayama, Meguro-ku, Tokyo, 152, Japan}$^{\dagger}$\\
\medskip{\it and}\\
\medskip{\it D.~V.~Vassilevich}\\
\medskip{\it Department of Theoretical Physics, St. Petersburg
University,}\\
{\it 198904 St.Petersburg, Russia}

\end{center}
\vspace{15mm}

\centerline{\bf Abstract}
\begin{quotation}
We derive the asymptotic expansion of the heat kernel for a Laplace
operator acting on deformed spheres. We calculate the
coefficients of the heat kernel expansion on  two- and
 three-dimensional deformed spheres as functions of  deformation
parameters. We find that
under some deformation the conformal anomaly for free scalar
fields on $R^4\times \tilde S^2$ and  $R^6\times \tilde S^2$ is canceled.

\end{quotation}

\vfill
\noindent
$^{\dagger}$ On leave from: {\em Irkutsk university, Russia};\\
Electronic mail: shtykov@phys.titech.ac.jp

\newpage

The asymptotic expansion of the heat kernel, corresponding to the
elliptic second-order differential operator acting on an
arbitrary manifold $M$ has been investigated in connection with
index theorems \cite{1} and some applications in field theory \cite{2,3}.
The kernel  $K(x,y,t)$  satisfies a heat equation
for the some second-order operator $H = -D^2 + X$ defined  on a
smooth N-dimensional Riemannian manifold ($X$ is a scalar function)
$$
(\partial_t\,+\,H)K(x,y,t) = 0
$$
with the boundary condition  $$K(x,y,0) = \delta(x,y).$$
The asymptotic expansion of $\ K(x,y,t)$ has been derived for
various models \cite{4}-\cite{7} in a general form \cite{6} and in
a numerical form for some homogeneous spaces \cite{7}. Under $t\RA 0$
the heat kernel has the following expansion
$$ K(x,y,t) = {(4\pi\,t)}^{-N/2}\,\Delta^{1/2}(x,y)\,
\exp{\,(-\sigma^2/4t)}\,\sum_{n=0}^{\infty}\,a_n(x,y)(t)^n
$$
where $\Delta$  is the invariant Van Vleck-Morette determinant
\cite{8}, $2\sigma(x,y)$ is the square of the geodesic distance
between $x$ and $y$. In terms of $ K(x,y,t)$ , one can write a simple
integral representation for the one-loop effective action . If one
takes regularization with the  short-distance cut-off  $L$ \cite{9}, the
regularized one-loop effective action $W^{(1)}$ can be defined as
$$
W^{(1)} =
\int^{\infty}_L\,\frac{dt}{t}\,K(t)
$$
Here $K(t) = tr\,\int\,d^Nxg^{1/2}\,K(x,x,t)$ with the asymptotic expansion
$$
K(t) = \sum_{n=0}^{\infty}\,A_n t^{n - N/2} = \sum_{n=0}^{\infty}\,
tr\,\int\,d^Nxg^{1/2}\,a_n(x,x)$$
The divergent terms in $W^{(1)}$ are proportional to the first
coefficients $a_n(x,x)$ . For even-dimensional spaces the most important
is the coefficient $a_{N/2}(x,x)$, since this single coefficient for
a given theory determines various anomalies \cite{10}.

In this letter we  explicitly calculate  the  coefficients $a_n(x,x)$ for
two- and three-dimensional
spaces obtained from the metric deformation of two- and
three-dimensional spheres respectively. We obtain the
coefficients $a_n$ as  functions of  the deformation
parameters and show that under some deformation the conformal
anomaly is canceled for free scalar fields defined on $\tilde
S^2\times R^4$ and $\tilde S^2\times R^6$.

Let us begin with the scalar Laplacian eigenvalues on deformed spheres.
The metric on the deformed sphere $\tilde S^{d+1}$ can be
expressed in the form
$$ ds^2=dx_0^2+\,sin^2x_0\,d\Omega^2 $$
where $d\Omega^2$ is the metric on the (deformed)
$\tilde S^d$. Any scalar function can be represented
as a sum of eigenfunctions $Y_{(l)}(x_i)$ of the
Laplace operator on $\tilde S^d$
\beq
\phi (x_0,x_i)=\sum_{(l)} f_{(l)}(x_0)
Y_{(l)}(x_i). \eeq
Substituting the decomposition (1) in the eigenvalue
equation
$$ \Delta \phi =\lambda \phi $$
we obtain the following ordinary differential
equation
\beq [\partial_0^2 +d {\rm ctg }x_0 \partial_0
-\frac {a_{(l)}}{\sin^2 x_0} ]f_{(l)}=\lambda f_{(l)}.
\eeq
The $-a_{(l)}$ is the eigenvalue of the Laplace
operator on $\tilde S^d$ corresponding to $Y_{(l)}$.
We shall drop the subscripts $(l)$ for a while.
Let us make the substitution
$$ f=h \sin^b (x_0), \quad b=\frac 12
(1-d+\sqrt {(1-d)^2+4a}) $$
and change the independent variable
$$z=\frac 12 (\cos x_0 +1). $$
The equation (2) then takes the form
$$z(z-1)h''+(1+c)(z-\frac 12 )h'+eh=0,$$
\beq e=b(b+d)+\lambda , c=2b+d. \eeq
Prime denotes  differentiation with respect to $z$.
According to the general prescription \cite{11} let us
express $h$ as the power series
\beq h(z)=\sum_{k=0} \alpha_k z^k.
\eeq
Substitution of (4) in (3) gives a recurrent condition
on the coefficients $\alpha_k$
\beq \alpha_{k+1}=\alpha_k
\frac {k(k-1)+(1+c)k+e}{(k+1)(k+(c+1)/2)}. \eeq
The denominator of (5) is positive for all $k$. The
eigenfunctins $h_k$ can be found by imposing the
condition on the numerator of (5) to be equal to
zero. We obtain  the eigenvalues
$$\lambda_{(l)k}=-k^2-(1+q)k-\frac 12 (1-d+q+2a_{(l)})$$
\beq q=\sqrt {(d-1)^2+4a_{(l)}} \eeq
where we restored the dependence on the index $(l)$. The
eigenvalues $a_{(l)}$ can be defined using the same
 formula (6) with $d \to d-1$. Repeating these steps
we can obtain the spectrum of scalar Laplace operator
on $\tilde S^{d+1}$ in terms of $d+1$ non-negative
integers and $d+1$ scale parameters.

For $d=3$ equation (6) was obtained in \cite{12} by the
same methods.

In the case of the unit round $d$-sphere $\tilde S^d$ with
$a_{(l)}=l(l+d-1)$ we obtain from (6)
$$ \lambda_{(l)k}=-(k+l)(k+l+d)=-n(n+d), \quad n=k+l
 $$
Thus  equation (6) reproduces the correct eigenvalues
of the scalar Laplace operator on the unit round
$S^{d+1}$. One can also verify that the degeneracies
have  the correct values.

 With the  deformation of a two-dimensional sphere, we consider rescaling
 $l^2\RA \rho l^2,\ \ (\rho >0)$ where $l^2$ are the eigenvalues
of a Laplace operator on the unit sphere $S^1$. The
eigenvalues (6) for $\tilde S^2$ can be written as
\beq
\lambda_{l,k} = - (k+ \rho l + 1/2)^2 + 1/4
\eeq
 The heat kernel for the
eigenvalues (7) is defined as
\beq
K(t) = K_1(t) +  K_2(t) = e^{t/4}\,(2 \sum_{l=1}^{\infty}\,
\sum_{k=\rho l + 1/2}^{\infty}\,e^{-k^2 t} +\,
\sum_{k=1/2}^{\infty}\,e^{-k^2 t})
\eeq
To derive the asymptotic expansion for the first term in (8)  we
rewrite the sum over $k$  by using the Mellin transform
$$
f(s,t) = \int_0^\infty\,dx\,x^{s-1}\,
e^{ -x^2t} = \frac{1}{2}\Gamma(\frac{s}{2})t^{-s/2}.$$
Performing the inverse transform
$$
\frac{1}{2\pi\,i}\,\int_{a-i\infty}^{a+i\infty}\,ds'\,k^{-s'}\,f(s',t)\,
$$
and summing over $k$  we obtain
\beq
 K_1(t) = e^{t/4}\,\frac{1}{2\pi\,i}\,\int_C\,ds'\,\sum_{l=1}^{\infty}\,
\Gamma (\frac{s'}{2})\,t^{-s'/2}\zeta(s', \rho l + 1/2) + R(t).
\eeq
Here the contour $C$ covers the poles of $\Gamma (\frac{s'}{2})$
at points $s'=-2m$ as well as
 poles of $\,g(s') = \sum_{l=1}^{\infty}\zeta(s', \rho l + 1/2)\,$ and
$$
R(t) = e^{t/4}\,\frac{1}{2\pi\,i}\,\int_D\,ds'\,\Gamma (\frac{s'}{2})
\,t^{-s'/2}\,g(s')
$$
where the contour $D$ consists of the semicircumference at infinity
on the left. The formula (9) is understood to be exact, but it is
difficult to compute $R(t)$ explicitly. However, one can show that
$R(t)$ vanishes exponentially as $t\RA 0$.
Thus, for small $t$, one can discard $R(t)$ relative to the power
series, leaving the asymptotic expansion for $K(t)$. (The calculations of
$R(t)$ for some series can be found in \cite{13}).
Using the Hermite formula \cite{11}
$$
\zeta(z,q) = \frac{q^{-z}}{2} + \frac{q^{1-z}}{z-1} + 2\int_{0}^{\infty}
dx\,\sin(z\,\arctan(x/q))\frac{(q^2 + x^2)^{-z/2}}{e^{2\pi x} - 1}
$$
for $\zeta(s', \rho l + 1/2)\,$ in (9), after summing over $l$ and
and integrating over $s'$ we obtain the following heat kernel expansion
$$
 K_1(t) = e^{t/4}\,(\frac{1}{\rho t} - \frac{\pi^{1/2}}{2t^{1/2}} + \sum_{m=0}^
{\infty}\,\frac{(-1)^m}{m!}\,\Bigl (-\frac{2}{2m+1}\rho^{2m+1}
\zeta(-2m-1,1+1/(2\rho))
$$
 \beq
+ \rho^{2m}\zeta(-2m,1+1/(2\rho))\, -\,\frac{2}{2m+1}\rho^{-2m-1}\zeta(-2m-1)
\,+\,F(-2m,\rho)\,\Bigr )
\eeq
where
$$
F(z,\rho)\,= \,2\sum_{p=0}^{\infty}\,(-1)^{p+1}c_p(z)\,\sum_{n=0}^{\infty}
\,\frac{\Gamma (n+z/2)}{\Gamma (z/2)n!}\rho^{-2p-2n-z-1}\, $$
 $$
\times \zeta(2p+2n+z+1,1+1/(2\rho))\zeta(-2p-2n-1),\ \ \ \ \ \ \ (2p+2n+z\neq
0\
).
$$
and the coefficients $c_p$ are determined from
$$
\sin(z\,\arctan(x))\, =\,\sum_{p=0}^{\infty}\,c_p(z)\,x^{2p+1}
$$
The asymptotic expansion for $K_2$ in (8) can be derived by using
the same method. After a little calculation (discarding the exponentially
small contribution) we find
  \beq
K_2(t)\,=\,e^{t/4}\frac{\pi^{1/2}}{2t^{1/2}},\ \ \ \ \ \ \ (t\RA 0).
\eeq
Substituting (10) and (11) in (8) and performing a numerical computation
we get
the following values of some $a_n(\rho)\ \ \ (a_0=1)$
$$ n\ \ \ \ \ \ \ \ \ \rho\,=\,0.2\ \ \ \ \ \ \ \ \ \ \ \ \rho\,=\,0.6
 \ \ \ \ \ \ \ \ \
\ \ \ \ \ \ \ \ \rho\,=\,1\ \ \ \ \ \ \ \ \ \ \ \ \ \ \ \rho\, =\,1.8$$
$$ 1\ \ \ \ \ \ \ \ \ 0.1733\ \ \ \ \ \ \ \ \ 0.2267\ \ \ \ \ \ \
\ \ \ \ \ \ 0.3333\ \ \ \ \ \ \ \ \ 0.7067 $$
 $$ 2\ \ \ \ \ \ \ \ \ 0.0077\ \ \ \ \ \ \ \ \ 0.0263\ \ \ \ \ \ \
\ \ \ \ \ \ 0.0667\ \ \ \ \ \ \ \ \ 0.2439 $$
$$ 3\ \ \ \ \ \ \ \ \ -0.0016\ \ \ \ \ \ \ \ \ 0.0024\ \
\ \ \ \ \ \ \ \ \ \ \ 0.0127\ \ \ \ \ \ \ \ \ 0.0902$$
\beq 4\ \ \ \ \ \ \ \ -0.0008\ \ \ \ \ \ \ \ \ 0.0003\ \ \ \ \ \ \
\ \ \ \ \ \ 0.0032 \ \ \ \ \ \ \ \ \ 0.0590
\eeq
For $\rho = 1$ we have from (12) in a numerical form the famous
asymptotic expansion for  unit round $S^2$
 $$ K(t) = \frac{1}{t} + 0.3333 + 0.0667 t + 0.0127 t^2 + 0.0032 t^3 + \ \ ...
$$

The next space we would like to consider is a three-sphere with another
homogeneous deformation which can be
represented as $ SU(2)\times U(1)/U(1) $ (the Taub space).
The eigenvalues of the
Laplace operator can be written as \cite{14}
\beq
\lambda_{n,j} = n^2 -1 + \omega(2j - n +1)^2
\eeq
where $\omega$ is the deformation parameter. The range of $\omega$
is $-1<\omega<\infty$ and $\omega = 0$ corresponds to round
$S^3$. Then the heat kernel takes the form
\beq
K(t) = \sum_{n=1}^{\infty}\,n\sum_{j=0}^{n-1}\,\exp{(-\,\lambda_{n,j})t}
\eeq
First we rewrite the sum over $j$ using the identity
$$
\sum_{j=0}^{n-1}\,\exp{(-\,\omega(2j - n +1)^2)t}\,=\,\Bigl (\sum_{
j=-(n-1)/2}^{\infty}\,-\,\sum_{(n+1)/2}^{\infty}\Bigr )e^{-4\omega j^2 t}
$$
Now it has the form similar to (8) and can be evaluated by means
 of the Mellin transform.
A straightforward calculation gives
\beq
K(t) = \,e^t\,\sum_{k=0}^{\infty}\frac{\omega^k(-1)^k(2k)!}{k!}
\sum_{r=0}^{2k}\frac{B_r\,2^r}{r!}\sum_{p=0}^{k-[(r+1)/2]}\,
\sum_{n=1}^{\infty}\frac{
e^{-n^2t}n^{2p+2}t^{k}}{(2k-2p-r)!(2p+1)!}
\eeq
Here we used the representation
$$ \zeta(-m, q) = - \sum_{r=0}^{m+1}\frac{m!B_rq^{m+1-r}}{r!(m-r+1)!}$$
where $B_r$ are Bernoulli numbers.
After similar manipulations with the sum over $n$ in (15) we obtain
$$
K(t) =
e^t\sum_{k=0}^{\infty}\frac{\omega^k(-1)^k(2k)!}{4k!}\sum_{r=0}^{2k}\frac{B_r\,\
2^r}{r!}\sum_{p=0}^{k-[(r+1)/2]}\frac{\Gamma(3/2+p)t^{k-p-3/2}}{(2k-2p-r)!(2p+1\
)!}
 $$
 $$
 = \frac{\pi^{1/2}}{4(1+\omega)^{1/2}}\Bigl ( 1 +\,
\frac{3+4\omega}{3(1+\omega\
)}$$
 \beq +  \frac{32\omega^2 + 40\omega
+15}{30(1+\omega)^{2}} + \,
\frac{369\omega^3 + 28\omega^2 + 140\omega +35}{210(1+\omega)^3}\,+\,...\Bigr )
\eeq
With $\omega = 0$ the expansion for round $S^3$ is reproduced.

As is known the divergencies in the one-loop effective action for
even-dimensional spaces lead
to scale symmetry breaking and give rise to a nonvanishing
conformal anomaly. The conformal anomaly has
a geometrical structure and is expressed by means of
$a_{N/2}$. In our case $a_n$ depend on the deformation parameters
and can be equal to zero with the appropriate parametric values .

Let us consider the one-loop effective action for scalar fields
on $R^m\times\tilde S^2$ where $R^m$ is Euclidean
$m$-dimensional space. The conformal anomaly arises when we take
the expectation value of the momentum-energy tensor
$T_{\mu}^{\mu}$ with the metric as a background classical field
$$ <T_{\mu}^{\mu}> = \frac{g_{\mu\nu}}{Z[g]}\,\frac{\delta
Z[g]}{\delta g_{\mu\nu}}$$
where $Z[g]$ is the generating functional of the theory.
Zeta-function regularization gives
\beq
<T_{\mu}^{\mu}> = \frac{1}{(4\pi)^{(m+2)/2}}\,a_{(m+2)/2}\,
\eeq
 From (10),(11),(8) one can compute that the
anomaly (17) for scalar fields on $R^4\times\,\tilde S^2$ and
$R^6\times\,\tilde S^2$ is removed with the values $\rho =0.41$ and
$\rho = 0.51$ respectivly. The Casimir energy is finite for these spaces
and can be computed explicitly. (Now this problem is under consideration).
For scalar fields on the 4-dimensional space $R^1\times\,SU(2)
\times U(1)/U(1)$  the
anomaly
$$
<T_{\mu}^{\mu}> = \frac{1}{(4\pi)^2}\frac{32\omega^2 +
40\omega +15}{30(1+\omega)^{2}}
$$
can not be removed with any value of $\omega$.

It should be noted that different type of deformed spheres have been
considered previously
in multidimensional models \cite{15}. However, only small one-parameter
deformations
have been used for the calculation of the one-loop potential. In our case
the deformation removing the conformal anomaly can not be considered small.\\

{\bf  Note added}\\

The manifolds with singular points were also studied in the context
of orbifold factors of spheres, and flat conical spaces.
The corresponding references can be found in the papers \cite{16},\cite{17}.
One of the us (D.V) is grateful to Guido Cognola for pointing
out Refs. \cite{17}.\\

{\bf Acknowledgements}

This work was partially supported by the Russian Foundation
for Fundamental Studies, grant 93-02-14378. One of us (DV)
is grateful to Ignati Grigentch for discussions on
Laplace operator on elliptic spaces.


\newpage

\begin{thebibliography}{99}

\bibitem{1}
Atiyah M, Bott R and Patodi V K 1973 Inv. Math {\bf 19} 279

\bibitem{2}
L$\ddot{u}$scher M 1982 Ann. Phys. {\bf 142} 359

\bibitem{3}
Leutwyler H 1985 Phys. Lett. {\bf 152 B} 78

\bibitem{4}
DeWitt B S 1975 Phys. Rep. {\bf 19} 295

\bibitem{5}
Barvinsky A O and Vilkovisky G A 1985 Phys. Rep. {\bf 119} 1

\bibitem{6}
Gilkey P B  1975 J. Diff. Geom. {\bf 110} 601; Jack I and
Parker L 1985 Phys. Rev. {\bf D31} 2439; Christensen S M 1976
Phys. Rev. {\bf D14} 2490; 1978 {\bf D17} 946; Avramidi I G 1989
Teor. Mat. Fiz.  {\bf 79} 219; Amsterdamski P, Berkin A L and
O'Connor D J 1989 Class. Quant. Grav.  {\bf 8} 1981

\bibitem{7}
Birmingham D 1987 Phys.  Rev. {\bf D36} 3037; Birmingham D,
Kantowski R and Milton K 1988 Phys. Rev. {\bf D38} 1069;
Lyakhovsky V D, Shtykov N N and Vassilevich D V  1991 Lett.
Math. Phys. {\bf 21} 89; Camporesi R 1990 Phys. Rep. {\bf 196} 1

\bibitem{8}
DeWitt Morette C, Makeshwari A and Melson B  1979 Phys. Rep.
{\bf 50} 255

\bibitem{9}
Schwinger J 1951 Phys. Rev. {\bf 82} 664

\bibitem{10}
Christensen S M and Duff M J  1979 Nucl. Phys. {\bf B154} 301;
Ball R and Osborn H  1986 Nucl. Phys. {\bf B263} 245

\bibitem{11}
Whittaker E T and Watson 1927 A course of modern analysis
(Cambridge: Cambridge Univ. Press)

\bibitem{12}
Grigentch I P and Vassilevich D V  1994 Nuovo Cimento {\bf 107A}
227

\bibitem{13}
Elizalde E and Romeo A  1989  Phys. Rev. {\bf D40} 436;
1989 J. Math. Phys. {\bf 30} 1133; Actor A 1991 J. Phys. A:
Math. Gen. {\bf 24} 3741

\bibitem{14}
Shen T C and Sobczyk  1987 Phys. Rev. {\bf D36} 397

\bibitem{15}
Lim C S 1985 Phys. Rev. {\bf D31} 2507

\bibitem{16}
Dowker J S 1990 in Formation and evolution of cosmic strings,
   eds. G.Gibbons et al, Cambridge Univ. Press; 1994 Class.
   Quantum Grav. {\bf 11} 557;
    Chang P and Dowker J S 1993 Nucl. Phys. {\bf B395 } 407

\bibitem{17}
Cognola G, Kirsten K and Vanzo L 1994 Phys. Rev. {\bf D49} 1027;
Cognola G and Vanzo L 1994 J. Math. Phys. {\bf 35}  3109;
Fursaev D V and Miele G 1994 Phys. Rev. {\bf D49} 987;
Fursaev D V 1994 Phys. Lett. {\bf B333} 326


\end{thebibliography}
\end{document}